\documentclass[12pt,a4paper]{article}
\usepackage{a4wide}
\usepackage{amssymb, amsmath}
\usepackage{graphicx}

\def\cH{{\cal{H}}}
\def\cD{{\cal{D}}}
\def\cU{{\cal{U}}}
\def\gR{\mathbb{R}}
\def\gT{\mathbb{T}}

\begin{document}

\title{Quantum Hall Conductivity in a Landau Type Model with a Realistic Geometry}
\author{F. Chandelier$^a$, Y. Georgelin$^a$, T. Masson$^b$, J.-C. Wallet$^a$}
\date{}
\maketitle
\begin{center}
$^a$ Groupe de Physique Th\'eorique,\\
Institut de Physique Nucl\'eaire\\
F-91406 Orsay Cedex, France\\
\bigskip
$^b$ Laboratoire de Physique Th\'eorique (UMR 8627)\\
B\^at 210, Universit\'e Paris-Sud Orsay\\
F-91405 Orsay Cedex
\end{center}

\bigskip
\begin{abstract}
In this paper, we revisit some quantum mechanical aspects related to the Quantum Hall Effect. We consider a Landau type model, paying a special attention to the experimental and geometrical features of Quantum Hall experiments. The resulting formalism is then used to compute explicitely the Hall conductivity from a Kubo formula.
\end{abstract}

\vfill
\begin{flushleft}
LPT-Orsay 02-113\\
IPNO DR 02-025
\end{flushleft}
\newpage

\section{Introduction}

It is well known that the integer values taken by the Hall conductivity can be explained and understood by using some topological arguments \cite{Laug:81, PranGirv:90, Ston:92}. In particular, it has been shown that the Kubo formula used to compute this Hall conductivity can be written into an explicit form in which the integral of the first Chern class of a certain line vector bundle appears \cite{NiuThouWu:85, Kohm:85, AvroSeil:85, NiuThou:87, Ston:92, BellElstScBa:94, Thou:94}. These topological arguments are relevant because one can introduce or recognize two extra parameters into the quantum mechanical description of the Quantum Hall Effect (QHE), which are subject to a specific periodic condition. Recall that the first indication in this direction was presented by Laughlin \cite{Laug:81} where however only one parameter (in the cylindrical geometry of \cite{Laug:81}) was involved. While the above topological based results are esthetically appealing, all the models presented so far have not been able to really predict the value of the Hall conductivity starting from some basic ingredients stemming from quantum mechanic. One reason for this may be due to the fact that the line vector bundle underlying those approaches did not take sufficiently into account the physical parameters of the experiment, that is the perpendicular applied magnetic field, the number of electrons involved in the phenomena, the shape of the sample, the nature of the threads, \dots 

Besides, many of the models for the QHE proposed so far have been based on some assumptions concerning the characteristics of the experiment which sometime do not fit quite well with the actual experimental situation. For instance, some of these models consider torus \cite{WenNiu:90} or  annular \cite{Halp:82} geometries for the two-dimensional sample. Let us note that these models consider the sample as well as the threads as quantum objects, if not simply ignoring the existence of the latter. Experimentally, such an assumption is certainly questionable.

The aim of this paper is to revisit a standard quantum mechanical model for the QHE while trying to incorporate as much as possible the specific geometrical and experimental constraints, and to derive from the resulting framework the Hall conductivity. The particular geometry of the QHE, already used in the literature \cite{NiuThou:87, AvroSeil:85}, and its physical consequences on the quantum mechanics of a two-dimensional interacting electrons gas, will be used as a guide to introduce the mathematical aspects of the model proposed here. Our computation of the Hall conductivity relies on the usual hypothesis on the validity of the Kubo formula.

In section~\ref{physicalmathematicalfeatures}, we recall the physical constraints of the experiment, and relate them to some mathematical properties of certain operators. In section~\ref{mathematicalmodel}, we introduce the mathematical formalism of our model. Some comments about this model are made in section~\ref{comments}. Section~\ref{computation} is devoted to the explicit computation of the Hall conductivity. In section~\ref{commentsandconclusion}, we conclude.

\section{Physical and mathematical features of the QHE}

\label{physicalmathematicalfeatures}

\begin{figure}
\begin{center}
\includegraphics{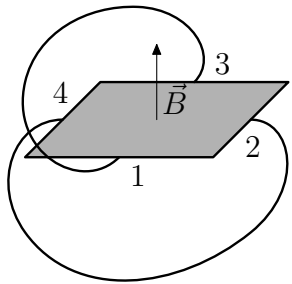}
\end{center}
\caption{Schematic representation of the global geometry of the QHE.}
\label{fig-general}
\end{figure}

Recall that the QHE takes place in a very specific physical situation where the electrons, maintained at extremely low temperature can be viewed as being confined to a two-dimensional finite domain on which a perpendicular magnetic field is applied. The electrons are injected and collected by wires at the four edges of the sample (see fig.~\ref{fig-general}). These wires are classical objects, connected to macroscopic devices (at room temperature) which, for instance, measure the intensity of the current. Quantum mechanics is required for the description of the electrons confined inside the sample, a rectangular planar domain as shown on fig.~\ref{fig-general}, while the electrons circulating in the wires must be regarded as classical objects. 

Keeping this in mind, the only two assumptions concerning the classical properties related to the wires we will make through this paper are:\\ 
i) the conservation of the currents circulating in the wires and\\ 
ii) the fact that the total charge located in one edge of the sample equals the one located in the opposited edge, so that the presence of the wires reflects itself as boundary conditions constraining the quantum mechanics ruling the electrons located inside the rectangular sample domain. 

Denoting by $j_i$, $i=1,2,3,4$ (see fig.~\ref{fig-general}) the outgoing signed current for the edge $i$, assumption i) can be written as
\begin{equation}
\label{currentconservation}
j_1 + j_3 = 0 \quad\text{ and }\quad j_2 + j_4 =0\ .
\end{equation}
These constraints can be reexpressed in term of integrals over each edge of the usual density current which appears in quantum mechanics in the Schr\"odinger representation. Note that these constraints are somehow different from some usual ({\sl i.e.} Neumann or Dirichlet) boundary conditions. Nevertheless, we will see in the following that they give rise naturally to specific mathematical properties for some operators on the Hilbert space of the problem.

A typical Quantum Hall experiment relies on two controlable (external) parameters, namely the applied perpendicular (strong) magnetic field $B$ and the number of electrons $N$ (number of charge carriers) inside the sample (which can be experimentally modified for instance by varying an applied gate voltage).

Most of the experimental data obtained from QHE experiments are measured in term of collective motions of the electrons, as are for example the currents in the wires and the Hall conductivity. Accordingly, we will be mainly interested into observables that can be expressed in terms of collective variables ({\sl i.e.} center of mass variables). Let us denote by $(x_i, y_i)$ and $(p_{x_i}, p_{y_i})$ ($i=1, \dots, N$) the individual position and momentum variables of each electron. Then the currents $j_k$ (in the absence of any magnetic field) depend only on the components of the total momentum $p_x = \sum_{i=1}^N p_{x_i}$ and $p_y = \sum_{i=1}^N p_{y_i}$ through the relation
\begin{subequations}
\label{currentdefinition}
\begin{equation}
j_1 = \int_{\text{edge 1}} j_y
\end{equation}  
where, as usual,
\begin{equation}
j_y = \psi^\ast (p_y \psi) + (p_y \psi)^\ast \psi
\end{equation}
\end{subequations}
together with similar relations for the $3$ other currents.

\bigskip
In the following, without lost of generality, we will consider a physical sample defined by the square $[0,1]^2 = [0,1] \times [0,1]$ in the $(x,y)$ plane. This slight assumption about the sample geometry will considerably simplify the discussion and most of the formulas involved in the ensuing discussion without altering the conclusions. We further assume from now on $\hbar=1$ and $e=1$ ($e$: electron charge).

Consider the ordinary Hamiltonian $H$ for $N$ identical spinless electrons of mass $m$ with mutual interactions in the $2$-dimensional square $[0,1]^2$, submitted to a constant perpendicular magnetic field $B$. $H$ is given by
\begin{equation}
\label{totalhamiltonian}
H = \sum_{i=1}^N \frac{1}{2 m}\left( (p_{x_i} + \frac{1}{2} B y_i)^2 + (p_{y_i} 
- \frac{1}{2} B x_i)^2\right) + \sum_{1\leq i<j \leq N} V(x_i - x_j, y_i - y_j)
\end{equation}
where $V$ is the interaction potential and the symmetric gauge for the vector potential has been used. Note that we could have chosen any linear gauge as we will point out in a while. This Hamiltonian must be a self-adjoint operator in a Hilbert space $\cH$, which looks like the $N$-fold tensor product of the Hilbert space $L^2([0,1]^2)$ of square integrable functions on $[0,1]^2$.

Keeping in mind the remark about collective variables given above, we choose to split the Hamiltonian into two parts, one part corresponding to the collective variables defined by $x = \frac{1}{N} \sum_{i=1}^N x_i, \  y = \frac{1}{N} \sum_{i=1}^N y_i, \   p_x = \sum_{i=1}^N p_{x_i}, \  p_y = \sum_{i=1}^N p_{y_i}$ and the other part involving ``internal'' variables defined by
\begin{equation}
\tilde{x}_i = x_i - x, \quad 
\tilde{y}_i = y_i - y, \quad 
\tilde{p}_{x_i} = p_{x_i} - \frac{1}{N} p_x, \quad  
\tilde{p}_{y_i} = p_{y_i} - \frac{1}{N} p_y
\end{equation}
for $i=1,\dots,N$ and satisfying the relations
\begin{equation}
\sum_{i=1}^N\tilde{x}_i = 0,\quad
\sum_{i=1}^N\tilde{y}_i = 0,\quad
\sum_{i=1}^N\tilde{p}_{x_i} = 0,\quad
\sum_{i=1}^N\tilde{p}_{y_i} = 0.
\end{equation}
We then obtain
\begin{subequations}
\begin{equation}
H = H_0 + H_I
\end{equation}
where
\begin{equation}
H_0 =  \frac{1}{2 N m}\left( (p_{x} + \frac{1}{2} N B y)^2 + (p_{y} - \frac{1}{2} N B x)^2\right)
\end{equation}
\end{subequations}
and $H_I$, the internal Hamiltonian, depends only of the variables $\tilde{x}_i, \tilde{y}_i, \tilde{p}_{x_i}$ and $\tilde{p}_{y_i}$ for $i=1, \dots , N-1$. The corresponding Hilbert space is then a tensor product 
\begin{equation}
\cH = \cH_0 \otimes \cH_I
\end{equation}
where the collective operators acts only on $\cH_0$ and the internal ones on $\cH_I$. Recall that this splitting is a standard procedure. 

Most of the ensuing analysis will be focused on the $H_0$ part of the Hamiltonian together with the corresponding Hilbert space $\cH_0$, since the observables we are interested in are operators acting only on $\cH_0$. This Hilbert space looks at first sight like $L^2([0,1]^2)$ for the variables $x,y$ but we will point out in the next section that this identification is not quite correct and must be considered more carefully.

Operators (especially unbounded ones) on some Hilbert space are not completely defined by their ``formal'' expressions. We have to specify the domain on which they are defined. This domain plays a central role in the operator theory, and it is known that in quantum mechanics a choice for such a domain in the case of an unbounded operator can be related to different physical situations. Consider as an example the ordinary quantum mechanics on the interval $[0,1]$. In this case, it is known that the self-adjointness of the unbounded momentum operator $p = -i \frac{\partial\hfill}{\partial x}$ is obtained by a choice of boundary conditions on the wave functions. These mathematical boundary conditions define a particular (dense) domain for $p$, and they can be interpreted as some particular physical restrictions on the boundaries. More precisely, one can show that there is a one parameter family of self-adjoint operators $p_\theta$ for $\theta \in [0,2 \pi[$ whose dense domain of definition is
\begin{equation*}
\cD_\theta = \{ \phi \in L^2([0,1]) /\ \phi \text{ absolutely continuous},\ \phi' 
\in L^2([0,1]),\ \phi(1) = e^{i\theta}\phi(0)  \}.
\end{equation*}
These operators are not unitary (and hence not physically) equivalent for different values of $\theta$.\footnote{A more instructive example is provided by the Aharonov-Bohm effect where it is well known that different self-adjoint extensions of the Hamiltonian correspond to different fluxes inside the solenoid.}

We now show that the boundary conditions that have been selected for the QHE at the beginning of this section can be used to choose a dense domain for the operators $D_x, D_y$ defined by
\begin{equation}
D_x = -i\frac{\partial\hfill}{\partial x} + \frac{1}{2} N B y \quad ; \quad D_y = -i\frac{\partial\hfill}{\partial y} - \frac{1}{2} N B x\ .
\end{equation}
Let us pick the wave functions $\psi$ and $\phi$ belonging to a sufficient regular class. Then, the following simple relations hold:
\begin{subequations}
\label{Pxrelation}
\begin{multline}
\label{Pxrelation1}
\int_0^1 \int_0^1(D_x \psi)^\ast(x,y) \phi(x,y) dx dy - \int_0^1 \int_0^1 
\psi^\ast(x,y) (D_x \phi)(x,y) dx dy = \\
 i \int_0^1 \left[ \psi^\ast (x,y) \phi(x,y) \right]_{x=0}^{x=1} dy
\end{multline}
\begin{multline}
\label{Pxrelation2}
\int_0^1 \int_0^1(D_x^2 \psi)^\ast(x,y) \phi(x,y) dx dy - \int_0^1 \int_0^1 
\psi^\ast(x,y) (D_x^2 \phi)(x,y) dx dy = \\
\int_0^1 \left[ \psi^\ast (x,y) (D_x\phi)(x,y) + (D_x \psi)^\ast(x,y) \phi(x,y) 
\right]_{x=0}^{x=1} dy
\end{multline}
\end{subequations}
together with similar relations for $D_y$. In (\ref{Pxrelation}) we use the notation $[f(x,y)]_{x=0}^{x=1} = f(1,y)-f(0,y)$. 

The left hand side of these relations can be more compactly written using the usual scalar product on the Hilbert space of square integrable functions. Namely, one obtains
\begin{subequations}
\begin{equation}
\label{Pxrelation1b}
(D_x \psi, \phi)-(\psi, D_x\phi) = i \int_0^1 \left[ \psi^\ast (x,y) \phi(x,y) \right]_{x=0}^{x=1} dy
\end{equation}
\begin{equation}
\label{Pxrelation2b}
(D_x^2 \psi, \phi)-(\psi, D_x^2\phi) = \int_0^1 \left[ \psi^\ast (x,y) (D_x\phi)(x,y) + (D_x \psi)^\ast(x,y) \phi(x,y) 
\right]_{x=0}^{x=1} dy\ .
\end{equation}
\end{subequations}

Now, the assumption ii) stating that there is no difference between the total charge located in one edge and in its opposite implies that the right hand side of (\ref{Pxrelation1b}) vanishes for $\psi = \phi$. Besides, assumption i) tells us that the currents associated with opposite side of the sample must satisfy relation (\ref{currentconservation}). Then, when $\psi = \phi$, the RHS of (\ref{Pxrelation2b}) can be written as $\int_0^1 \left[ j_x(x,y) \right]_{x=0}^{x=1} dy$ where $j_x(x,y) = \phi^\ast (x,y) (D_x\phi)(x,y) + (D_x \phi)^\ast(x,y) \phi(x,y)$ is the usual expression for the current in the $x$-direction for an applied magnetic field $B$ so that assumption i) implies that the RHS of (\ref{Pxrelation2b}) vanishes. Summarizing the above considerations, any (sufficiently regular) wave function $\phi$ satisfying the boundary conditions i) and ii) obeys the following relations
\begin{subequations}
\begin{align}
(D_x \phi, \phi)-(\phi, D_x\phi) & = 0  & (D_y \phi, \phi)-(\phi, D_y\phi) & = 0 
\\
(D_x^2 \phi, \phi)-(\phi, D_x^2\phi) &= 0 & (D_y^2 \phi, \phi)-(\phi, D_y^2\phi) 
&= 0\ .
\end{align}
\end{subequations}

Let us denote by $\cD$ the linear space of states compatible with the two boundary conditions. In other words, if  $\psi,\phi \in \cD$, then any linear combinaison of $\psi$ and $\phi$ belongs to $\cD$. Using polarization formulas, it is easy to show that the relations
\begin{subequations}
\label{PxyP2xyselfadjoint}
\begin{align}
(D_x \psi, \phi)-(\psi, D_x\phi) & = 0  & (D_y \psi, \phi)-(\psi, D_y\phi) & = 0 
\label{Pxyselfadjoint}\\
(D_x^2 \psi, \phi)-(\psi, D_x^2\phi) &= 0 & (D_y^2 \psi, \phi)-(\psi, D_y^2\phi) 
&= 0 \label{P2xyselfadjoint}
\end{align}
\end{subequations}
hold also for any $\psi,\phi \in \cD$. So, $\cD$ is the linear space of functions on $[0,1]^2$ stable by the operators $D_x$ and $D_y$, and on which (\ref{PxyP2xyselfadjoint}) are satisfied. These relations clearly imply that $D_x, D_y, D_x^2$ and $D_y^2$ are self-adjoint operators on $\cD$.

So we have translated our physical boundary conditions imposed by the geometry and the physics of the QHE into a mathematical formulation of self-adjointness about operators on the Hilbert space involved in the problem. Now we are going to take these mathematical conditions as the starting point for the construction of the mathematical framework that will permit us to calculate the Hall conductivity.

\section{The mathematical framework}

\label{mathematicalmodel}

Let us introduce the operator algebra for the operators whose action must be represented on the Hilbert space $\cH_0$. These latter are the two coordinates operators associated with the variables $(x,y)$, that we denote by $X$ and $Y$. The operators $D_x$ and $D_y$ are the ``differential expressions'' for the operators $p_x - A_x$ and $p_y - A_y$ respectively, where the curl of the potential vector $(A_x, A_y)$ is $N$ times the applied magnetic field $B$. The appearance of the quantity $NB$ is related to the fact that we work with the center of mass variables $(x,y)$. We denote by $P_x$ and $P_y$ these operators. The four operators $X,Y,P_x, P_y$ must obey the following constraints. They obey to the commutations relations:
\begin{align}
[X,Y] &= 0 & [P_x, P_y] & = i NB\nonumber \\
[P_x, X] &= -i & [P_x, Y] &=0 
\label{algebra}\\
[P_y, X] &= 0 & [P_y, Y] &= -i\ .\nonumber
\end{align}
Besides, the spectrum of $X$ and $Y$ is the unit interval $[0,1]$. Finally, $P_x, P_y, P_x^2$ and $P_y^2$ are self-adjoint.

The commutations relations can be obtained using the ordinary commutation relations between the operators $x_i, y_i, p_{x_i}$ and $p_{y_i}$ in the presence of the constant magnetic field while the spectrum of the operators $X$ and $Y$ is just related to the finite size of the physical sample. Note that the self-adjointness of $P_x, P_y, P_x^2$ and $P_y^2$ is a consequence of the discussion about boundary conditions. 

Consider now the Hamiltonian $H_0$ which can be expressed as
\begin{equation}
\label{hamiltonian0}
H_0 = \frac{1}{2 N m} ( P_x^2 + P_y^2)
\end{equation}

Contrary to the one dimensional situation, there is no general method for representing $P_x$ and $P_y$ as self-adjoint operators. Here, we will make an Ansatz, inspired by the one dimensional situation recalled above. Indeed, one knows that $P_x$ and $P_y$ must behave like the first order differential operators $D_x$ and $D_y$. Keeping this in mind, we shall take the following Ansatz for the representation of the algebra (\ref{algebra}):
\begin{enumerate}
\item The domain on which $P_x$ and $P_y$ are defined is the linear space of functions $\phi \in L^2([0,1]^2)$, absolutely continuous in the two variables, whose first derivatives belong to $L^2([0,1]^2)$. The boundary conditions are of the form
\begin{equation}
\label{ansatzboundary}
\phi(1,y) = e^{-i f(y)} \phi(0,y) \quad ; \quad 
\phi(x,1) = e^{-i g(x)} \phi(x,0)
\end{equation}
for any real functions $f,g$. These relations can be viewed as a generalization of the one dimensional case (see the definition of $\cD_\theta$).
\item The operators $P_x$ and $P_y$ are first order differential operators acting on these functions, whose expression are
\begin{equation}
\label{ansatzPxPy}
P_x = -i\frac{\partial\hfill}{\partial x} + a(y) \quad ; \quad 
P_y = -i\frac{\partial\hfill}{\partial y} + b(x)
\end{equation}
for two functions $a$ and $b$.
\end{enumerate}
The commutation relations between $P_x$ and $P_y$ then imply
\begin{equation}
\frac{\partial a(y)}{\partial y} = NB + \frac{\partial b(x)}{\partial x}\ .
\end{equation}
One easily observes that the right (resp. left) hand side of this latter expression depends only on $x$ (resp. $y$), so that it must be a real constant $C$. The functions $a$ and $b$ are then of the form
\begin{equation}
a(y) = C y - \gamma \quad \text{ and }\quad 
b(x) = (C - NB)x - \eta 
\end{equation}
for two real constants $\gamma, \eta$. From this, it follows that $P_x = -i\frac{\partial\hfill}{\partial x} + C y - \gamma$ and $P_y = -i\frac{\partial\hfill}{\partial y} + (C - NB)x - \eta $. Further performing a gauge transformation given by $\phi(x,y) \mapsto e^{-i(\gamma x + \eta y)} \phi(x,y)$, one can get rid of the constants $\gamma, \eta$ appearing in the explicit expressions for $P_x$ and $P_y$. These constants will however reappear as arbitrary constants in the definitions of $f$ and $g$ (see (\ref{ansatzboundary})). Let us define $C= -\beta NB$ and $C- NB= \alpha NB$, with $\alpha$ and $\beta$ satisfying the relation 
\begin{equation}
\alpha + \beta = -1\ .
\end{equation}
The operators $P_x$ and $P_y$ then take the form
\begin{equation}
\label{PxandPy}
P_x = -i\frac{\partial\hfill}{\partial x} - \beta NB y \quad ; \quad
P_y = -i\frac{\partial\hfill}{\partial y} + \alpha NB x\ .
\end{equation}
At this level, one comment is in order. The arbitrariness in the choice of $\alpha$ (or $\beta$) is related to a gauge fixing for the vector potential whose curl is $NB$. For instance, when $\alpha=\beta = -1/2$, $P_x$ and $P_y$ coincide with $D_x$ and $D_y$. In the following, we will not fix any particular value for $\alpha, \beta$. Furthermore, it can be easily shown, by gauge transformation, that taking more general dependence of the functions $a$ and $b$ on the variables $x$ and $y$ would simply means considering a unitary equivalent representation of our algebra (\ref{algebra}).

One observes now that the operators $P_x$ and $P_y$ have a form similar to the one for some covariant derivatives on a line vector bundle over a two dimensional torus $(x,y) \in\gT^2$. Note however that it is not the case for arbitrary value of $NB$, as we will point out in a while. From (\ref{PxandPy}), it is easy to deduce the possible boundary conditions of the form (\ref{ansatzboundary}). They are given by
\begin{equation}
\phi(1,y) = e^{i \gamma - i \alpha NB y} \phi(0,y) \quad ; \quad 
\phi(x,1) = e^{i \eta + i \beta NBx} \phi(x,0)\ .
\end{equation}
The two real parameters $\gamma, \eta$ are arbitrary and they label inequivalent representations of $P_x, P_y$ for different values (modulo $2 \pi$). We will denote by $\cD_{\gamma,\eta}$ the domain on which are defined the $P_x$ and $P_y$ that we have constructed just above for specific values of $(\gamma, \eta)$. It appears as we will see in the sequel that it is very convenient to keep the same notation $P_x, P_y$ for the different representations of these operators, so that the $(\gamma, \eta)$ dependence will not be explicitely indicated. As already mentionned, for different values of $(\gamma, \eta)$ which are equal modulo $2 \pi$, the representations are equivalent. So we take as fondamental domain for $(\gamma, \eta)$ the square $[0, 2\pi]^2$. Nevertheless, we keep in mind that it will be very convenient for further considerations to allow for all possible real values for $(\gamma, \eta)$. The general structure in the $(\gamma, \eta)$-plane will be explored in a while.

We emphasize that there is neither physical nor mathematical constraints on the parameters $(\gamma, \eta)\in [0, 2\pi]^2$. Actually, from a physical viewpoint, it is very natural to consider all their possible values. This reflects the fact that considering at once a large number of possible mathematical boundary conditions corresponds to a unique physical state for the classical current through the threads. This can be translated on a more mathematical footing by considering the relevant Hilbert space appearing in the problem equal to the direct sum of the Hilbert spaces associated to each individual values of the couple $(\gamma, \eta)$. This direct sum takes the form of a direct hilbertian integral over $[0, 2\pi]^2$ of isomorphic Hilbert spaces (each representation being defined in the same Hilbert space $L^2([0,1]^2)$). A general wave function is then an element of $L^2([0,1]^2\times [0, 2\pi]^2)$. Let us introduce in this Hilbert space the linear subspace $\widetilde{\cD}$ of functions $\psi(x,y,\gamma,\eta)$ which are continuous in $(\gamma, \eta)$, and such that for any pair $(\gamma, \eta)$, $\psi(x,y,\gamma,\eta)$ is an element of $\cD_{\gamma,\eta}$. These functions then satisfy the following boundary relations
\begin{equation}
\label{boundarypsi}
\psi(1,y,\gamma,\eta) = e^{i \gamma - i \alpha NB y} \psi(0,y,\gamma,\eta) 
\quad ; \quad
\psi(x,1,\gamma,\eta) = e^{i \eta + i \beta NBx} \psi(x,0,\gamma,\eta)\ .
\end{equation}
When $NB$ is different from $2 \pi \ell$ for any integer $\ell$, we further impose that these functions vanish at the four points $(x,y)=(0,0)$, $(0,1)$, $(1,0)$ and $(1,1)$, which is dictated by (\ref{boundarypsi}). The space $\widetilde{\cD}$ is chosen to be the domain on which $P_x$ and $P_y$ are defined.

We choose a partial normalization on the functions $\psi$ of the form
\begin{equation}
\label{partialnormalizationpsi}
\int_{[0,1]^2} dx dy (\psi^\ast \psi)(x,y,\gamma,\eta) = 1
\end{equation}
for all $\gamma,\eta$ so that $\psi$ is normalized to unity, namely
\begin{equation}
\frac{1}{(2\pi)^2} \int_{[0,2\pi]^2} d\gamma d\eta \int_{[0,1]^2} dx dy (\psi^\ast \psi)(x,y,\gamma,\eta) = 1\ .
\end{equation}
This integral completely defines the scalar product on our Hilbert space.

Consider now the operators $X,Y$. We can represent them as multiplicative operators by the coordinates $x$ and $y$ respectively on a dense domain of functions which vanish on the edges of the square $[0,1]^2$. The spectrums of $X$ and $Y$ are then $[0,1]$. If one needs to work with $X,Y,P_x,P_y$ at the same time, one must consider the intersections of the domain $\widetilde{\cD}$ and the domain on which $X,Y$ are defined. This reduces considerably the interesting structure for $\widetilde{\cD}$, because functions in $\widetilde{\cD}$ must then vanish at the boundary of the square $[0,1]^2$, eliminating any dependence on $\gamma$ and $\eta$. Fortunately, we will not have to deal with all these operators at the same time. Actually, only the operators $P_x$ and $P_y$ will be used. In particular, we have seen that the Hamiltonian does not depend on $X$ and $Y$. So in the following, we will focalize ourselves on the structure of $\widetilde{\cD}$. 

One can now consider a dense domain $\cD \subset L^2([0,1]^2\times [0, 2\pi]^2)$ for the operators $P_x^2$ and $P_y^2$. It is the restriction of the domain $\widetilde{\cD}$ defined by the condition that $\psi$ belongs to $\cD$ if $\psi$, $P_x \psi$ and $P_y \psi$ are in $\widetilde{\cD}$. In particular, the functions in this domain satisfy the relations (\ref{boundarypsi}) and the Hamiltonian $H_0$ (\ref{hamiltonian0}) is well defined on $\cD$.

A straightforward computation shows that $P_x$ and $P_y$ (resp. $P_x^2$ and $P_y^2$) are self-adjoint operators on the dense domain $\widetilde{\cD}$ (resp. $\cD$), so that all wave functions in $\cD$ satisfy our physical boundary conditions (\ref{PxyP2xyselfadjoint}). In the rest of this paper, the dense domain relevant for the problem will be identified with $\cD$.

\section{Discussion of the mathematical framework}

\label{comments}

We now make some comments about the mathematical framework we have constructed in the previous section. 

First, notice that we did not prove the uniqueness of this representation. However, the Ansatz we have chosen gives us enough generality to obtain this rich structure. Moreover, this representation is not irreducible. This is an important feature, which is related to the full generality of the boundary conditions we have imposed. As already mentionned, this is actually necessary in order to take into account the largest possible number of mathematical boundary conditions compatible with our physical conditions. Besides, performing an arbitrary gauge transformation on this representation does not alter its corresponding main structures. Note that the  arbitrariness on the parameters $\alpha$ and $\beta$ already take into account a large class of gauge choices.

The boundary conditions (\ref{boundarypsi}) are very similar to some relations that the sections of a line vector bundle must satisfy. Indeed, we could be tempted to consider the space $(x,y) \in [0,1]^2$ as a two dimensional torus, and the functions $\psi \in \cD$ as sections of a line vector bundle over it. But this is \emph{not possible} for all values of $NB$, as it can be easily verified by considering the variations of the functions along a path on the edge of the square $[0,1]^2$. As already briefly noticed, the particular values $NB = 2 \pi \ell$ (for an integer $\ell$) are singled out in this respect. Only when $NB = 2 \pi \ell$ is it possible to consider the functions $\psi \in \cD$ as sections of a line vector bundle over a two dimensional torus. The condition $NB = 2 \pi \ell$ reflects the known fact that the first Chern class of such a bundle must be an integer. Notice that this underlying torus does not actually exist. In the following, we will study the structure of $\cD$ assuming that $NB = 2 \pi \ell$.

If one performs the gauge transformation 
\begin{equation}
\label{gaugetransformation}
\widetilde{\psi}(x,y,\gamma,\eta) = e^{-i(\gamma x + \eta y)} \psi(x,y,\gamma,\eta)
\end{equation}
then the operators $P_x$ and $P_y$ become $\widetilde{P}_x = P_x + \gamma$ and $\widetilde{P}_y = P_y + \eta$.  The boundary conditions for $\widetilde{\psi}$ simplify to
\begin{equation}
\label{boundarypsitilde}
\widetilde{\psi}(1,y,\gamma,\eta) = e^{- i \alpha NB y} \widetilde{\psi}(0,y,\gamma,\eta) 
\quad ; \quad
\widetilde{\psi}(x,1,\gamma,\eta) = e^{i \beta NBx} \widetilde{\psi}(x,0,\gamma,\eta)\ .
\end{equation}
As a counterpart, the $(\gamma, \eta)$ dependence is completely transfered into the operators $\widetilde{P}_x$ and $\widetilde{P}_y$. Obviously, both representations give rise to the same physics, and we will make use of this second representation in the next section to compute the Kubo formula. 

We have already noticed that the indices $(\gamma, \eta)$ of these different inequivalent representations need not to vary in $\gR$ but only in $[0, 2\pi]$ because if, for instance, $\gamma = 2\pi$ we get a represention which is equivalent to the one for $\gamma = 0$. Even when the representations are equivalent, the functions $(x,y,\gamma,\eta) \mapsto \psi(x,y,\gamma,\eta)$ and $(x,y,\gamma,\eta) \mapsto \psi(x,y,2\pi + \gamma,\eta)$ are not the same. Actually, the space of the latter is unitarily equivalent to the space $\cD$ of the former. We will not use explicitely this unitary correspondence. In the next section, we will use the functions $\psi(x,y,\gamma,\eta)$ \emph{as if} they were defined for $(\gamma, \eta) \in \gR^2$. This will be a convenient way to write some relations between them, and it will have to be understood modulo unitary correspondences.

If one applies a small electric field for instance in the $x$ direction, this electric field contributes only in $H_0$, because each electron undergoes the same influence. Indeed, in the total Hamiltonian, one adds the sum $\sum_{i=1}^N E x_i$ to $H$. This term depends only on $x$, and so must be a part of $H_0$. One alternative way to implement such an electric field is to create an increasing magnetic flux in one of the loops that the wires form in the experiment (see fig.~\ref{fig-general}). This (very slowly) increasing flux $\phi(t)$ will then comes out as an extra additional contribution to one of the operators $P_x$ or $P_y$. The resulting currents induced by this electric field can then be computed by the linear response formula of Kubo \cite{Kubo:57, Imry:97}, for the total Hamiltonian $H$. In particular, the Hall conductivity can be obtained in this way. Notice that because this electric field appears only in $H_0$, the Kubo formula will factor out the contribution from $H_I$. 

Now, it is known that this Kubo formula can be expressed in term of the first Chern class of a line vector bundle \cite{AvroSeilSimo:83, NiuThouWu:85, Kohm:85, AvroSeil:85, NiuThou:87, BellElstScBa:94}. The key procedure to do this is to express the velocity operators in terms of some derivatives of the Hamiltonian operator with respect to some variables. These variables define the base space of the line vector bundle mentioned before. We refer to the literature for the exact expressions, and to Appendix~\ref{kuboformula} for the actual formula we have to compute in our framework. In the present case, the natural variables are the pair $(\gamma, \eta)$ when one takes the new representation (\ref{gaugetransformation}). Indeed, as already noticed, the $(\gamma, \eta)$ counterparts appear as additional terms in the momentum operators. The derivatives of the corresponding Hamiltonian $\widetilde{H}_0$ with respect to $\gamma$ and $\eta$ gives then the velocities in the $x$ and $y$ directions. We assume in what follows that the Kubo formula can be applied in the present situation. In particular, this requires the occurrence of some properties concerning the Fermi gap and some extra conditions about the degeneracy of the ground-state which will appear later (see Appendix~\ref{kuboformula}). These conditions apply to the total Hamiltonian $H$ (\ref{totalhamiltonian}).

Clearly, the approach to the QHE we propose here bears some similarities with others approaches that have been reported in the past, in particular the appearance of two extra parameters defined on a torus, a square or a rectangle ($(\gamma, \eta)$ in our case). We would like to mention some of them. 

The first one can be found in Thouless {\sl et al.} \cite{ThouKohmNighNijs:82} and Kohmoto \cite{Kohm:85}. In these papers, the two extra parameters are related to the underlying crystallographic structure of the sample. They are generalized crystal momenta, which parameterize the Bloch waves functions, and are restricted in a magnetic Brillouin zone, which is a finite rectangle. The Kubo formula is then related to the first Chern class of a line vector bundle over this Brillouin zone. In this approach, the underlying torus is connected to some microscopic periodicity. This kind of model has the drawback that one has to assume an infinite sample. However, these pionneering investigations have been a starting point for further developpements. In particular, all the essential features coming from the presence of these two extra parameters have been exposed.

The second kind of approach we would like to comment can be found in Niu and Thouless \cite{NiuThou:87} and Avron and Seiler \cite{AvroSeil:85}. In this approach, the global geometry of the QHE of fig.~\ref{fig-general} is used to introduce the extra parameters. They are connected to the two fluxes through the loops of the wires. One of them is used to impose an electric field in the sample (as already mentioned), and the other is used to measure the response of the system. By a gauge transformation, it is easy to see that these fluxes have periodic effects on the system, and are therefore restricted to the two dimensional torus $[0,2\pi]^2$. The salient feature of this approach is that it does not suppose any specific microscopic properties such as a crystallographic periodicity. Notice that the use of some periodic flux of a magnetic field (not connected to the magnetic field in the sample) goes back to the Laughlin's argument \cite{Laug:81}, in which a cylindrical geometry was considered. 

Finally, our model bears some similarity to the one proposed by Niu, Thouless and Wu \cite{NiuThouWu:85}. In this model, the extra parameters are some phases in the boundary conditions of the wave functions over a rectangular sample. Although they mention the hermiticity of the Hamiltonian as a requirement for their choice of these boundary conditions, they do not strongly relate them to some physical experimental origins as we did. However, except for instance the separation of the variables into the global and the ``internal'' ones, some of the essential features of our model are already involved in \cite{NiuThouWu:85}.

We must stress that our approach involves the representation of an abstract algebra by operators (with specific adjointness properties) on a Hilbert space. Some other propositions has been made in this direction previously. One of them has been performed by Gr\"umm, Narnhofer and Thirring \cite{GrumNarnThir:85} and Gr\"umm \cite{Grum:85}. These authors consider an abstract algebra generated by the position and momentum operators similar to the one we introduce, and represent it on a Hilbert space for different underlying geometries. Unfortunately, these geometries are not enough realistic to produce a pertinent model for integral and \emph{fractional} QHE.

\section{Computation of the conductivity}

\label{computation}

All the models mentionned in the previous section give rise to a Hall conductivity that takes integer or fractional values (in units $e^2/h$). As far as we know, no one has permitted to \emph{compute} it, and to relate it to the filling factor $\nu$. In this section, we show that when the product $NB$ satisfy the integrity condition $NB= 2\pi\ell$, this computation can be performed entirely.

Recall that when $NB= 2\pi\ell$ for an integer $\ell$, one can take advantage of the properties of a line vector bundle over the two dimensional torus with periodicity $1$ in the variables $(x,y)$. This permits us to write $\psi(x,y,\gamma,\eta)$ for values of $x$ and $y$ greater than $1$. In particular, one has
\begin{align}
\label{boundarypsitorus1}
\psi(x+1,y,\gamma,\eta) &= e^{i \gamma - i \alpha NB y} \psi(x,y,\gamma,\eta) \\
\label{boundarypsitorus2}
\psi(x,y+1,\gamma,\eta) &= e^{i \eta + i \beta NBx} \psi(x,y,\gamma,\eta)\ .
\end{align}
Besides, as previously discussed, we consider the variables $\gamma$ and $\eta$ belonging to $\gR$. In order to compute the Kubo formula, one has to compute the first Chern class of the line vector bundle over $(\gamma, \eta) \in [0,2\pi]^2$. Let us characterize this line vector bundle. One way to do that is to look for the possible symmetries of the system. Such a symmetry has to commute with $H_0$, to act only on the collective variables (so is a symmetry of $\cH_0$) and to leave globally invariant the dense domain $\cD$. It appears that there exists two commuting unitary operators having these properties. Indeed, if one looks for some unitary operator of the form 
\begin{equation}
\psi(x,y,\gamma, \eta) \mapsto e^{i(a x + b y)} \psi(x + c, y + d, \gamma + e, \eta + f)
\end{equation}
which commutes with $H_0$ and leaves $\cD$ invariant (which is equivalent to leave invariant the relations (\ref{boundarypsitorus1}) and (\ref{boundarypsitorus2})), then one obtains a two parameters family of unitary operators $\cU_{\theta, \sigma}$ defined by
\begin{equation}
\label{twoparametersfamily}
(\cU_{\theta, \sigma} \psi)(x,y,\gamma, \eta) = e^{i(\beta NB \sigma x + \alpha NB \theta y)} \psi(x + \theta, y - \sigma, \gamma + NB\sigma, \eta + NB\theta)\ .
\end{equation}
Actually, a straightforward computation shows that these operators commute not only with $H_0$, but also with $P_x$ and $P_y$. So, one has $\cU_{\theta, \sigma} \cD = \cD$, and $[P_x, \cU_{\theta, \sigma}] = [P_y, \cU_{\theta, \sigma}] = 0$. But for different couples of the parameters $(\theta, \sigma)$ these operators do not commute. Actually, one has
\begin{equation}
\cU_{\theta, \sigma} \cU_{\theta', \sigma'} = e^{-iNB(\sigma' \theta - \sigma \theta)} \cU_{\theta', \sigma'} \cU_{\theta, \sigma}\ .
\end{equation}
Note that these unitary operators could be considered as some generalization of the usual magnetic translation operators.

As explained previously and further developed in Appendix~\ref{kuboformula}, the Hall conductivity can be computed using the gauge transformed functions $\widetilde{\psi}$ defined by (\ref{gaugetransformation}) in the Kubo formula. In the following, we will work with these functions. The $2$-parameter family of unitary operators (\ref{twoparametersfamily}) takes then the form
\begin{equation}
\label{twoparametersfamilytransformed}
(\widetilde{\cU}_{\theta, \sigma} \widetilde{\psi})(x,y,\gamma, \eta) = e^{-i(\alpha NB \sigma x - \beta NB \theta y - \gamma \theta + \eta \sigma)} \widetilde{\psi}(x + \theta, y - \sigma, \gamma + NB\sigma, \eta + NB\theta)\ .
\end{equation}
We take the two special values corresponding to the $2\pi$ translation in the variables $\gamma$ and $\eta$ :
\begin{equation}
\widetilde{U}^\frac{1}{\ell} = \widetilde{\cU}_{1/\ell, 0} \quad ; \quad \widetilde{V}^\frac{1}{\ell} = \widetilde{\cU}_{0, 1/\ell}\ .
\end{equation}
These operators satisfy to the commutation relations
\begin{equation}
\label{commutationrelations}
\widetilde{V}^\frac{1}{\ell} \widetilde{U}^\frac{1}{\ell} = e^{i \frac{2\pi}{\ell}} \widetilde{U}^\frac{1}{\ell} \widetilde{V}^\frac{1}{\ell}
\end{equation}
so that the two operators $\widetilde{U} = \left( \widetilde{U}^\frac{1}{\ell} \right)^\ell$ and $\widetilde{V} = \left( \widetilde{V}^\frac{1}{\ell} \right)^\ell$ commute with $\widetilde{U}^\frac{1}{\ell}$ and $\widetilde{V}^\frac{1}{\ell}$.

One knows that unitary operators satisfying (\ref{commutationrelations}) can be represented as $\ell\times \ell$ matrices of the form
\begin{equation}
\widetilde{U}^\frac{1}{\ell} = e^{i\frac{u}{\ell}}
\begin{pmatrix}
1 & 0 & \dots & 0 \\
0 & q & \dots & 0 \\
\hdotsfor{4} \\
0 & 0 & \dots & q^{\ell -1} 
\end{pmatrix} 
\quad ;\quad 
\widetilde{V}^\frac{1}{\ell} = e^{i\frac{v}{\ell}}
\begin{pmatrix}
0 & 1 & 0 & \dots & 0 \\
0 & 0 & 1 & \dots & 0 \\
\hdotsfor{5} \\
0 & 0 & 0 & \dots & 1 \\
1 & 0 & 0 & \dots & 0 
\end{pmatrix} 
\end{equation}
where $q = e^{i \frac{2\pi}{\ell}}$ and $u$ and $v$ are real parameters. The operators $\widetilde{U}$ and $\widetilde{V}$ therefore represent multiplication by respectively $e^{i u}$ and $e^{i v}$. Our Hilbert space can then be decomposed into direct sums of such representations (sectors), labelled by $(u,v)$ up to some extra parameters (extra degeneracy). So, one can introduce $\ell$-uplets of orthogonal functions\footnote{Notice that the rigged Hilbert spaces formalism should be used here in order to take into account a possible continuous part for the spectrum of $\widetilde{U}$ and $\widetilde{V}$. Here, we don't use explicitely this formalism, but we nevertheless keep it in mind in order to make meaningfull some expression appearing in the text.} $(\widetilde{\psi}_i)_{i=1, \dots, \ell}$ which satisfy, if one uses the explicit expression for $\widetilde{U}^\frac{1}{\ell}$ and $\widetilde{V}^\frac{1}{\ell}$
\begin{align}
\label{uaction}
\widetilde{\psi}_i(x,y,\gamma,\eta) &= q^{i-1} e^{-i(\frac{u}{\ell} + 2 \pi \beta y - \frac{\gamma}{\ell} )} \widetilde{\psi}_i(x + \frac{1}{\ell},y,\gamma,\eta + 2 \pi) \\
\label{vaction}
\widetilde{\psi}_{i+1}(x,y,\gamma,\eta) &= e^{-i(\frac{v}{\ell} + 2 \pi \alpha x - \frac{\eta}{\ell} )} \widetilde{\psi}_i(x,y - \frac{1}{\ell},\gamma + 2 \pi,\eta)\ . 
\end{align}
The relation (\ref{uaction}) tells us that the function $\widetilde{\psi}_i$ restricted to $\eta \in [2 n\pi, 2 (n+1)\pi]$, for $1 \leq n \leq \ell -1$, is completely determined by its restriction on $\eta \in [0, 2\pi]$. Similarly, relation (\ref{vaction}) means that the function $\widetilde{\psi}_{i+1}$ restricted to $\gamma \in [2 m\pi, 2 (m+1)\pi]$ can be obtained by the restriction of $\widetilde{\psi}_{i}$ on $\gamma \in [2 (m-1)\pi, 2 m\pi]$ for $1 \leq m \leq \ell -1$. Globally, the functions $\widetilde{\psi}_i$ are completely characterized either by their $\ell$ restrictions to the elementary cell $(\gamma, \eta) \in [0, 2\pi]^2$ or by only one of them restricted to $(\gamma, \eta) \in [0, 2\pi\ell]\times [0, 2\pi]$. Applying the operators $\widetilde{U}$ and $\widetilde{V}$, and using the boundary relations (\ref{boundarypsitorus1}) and (\ref{boundarypsitorus2}), one can show that each function $\widetilde{\psi}_{i}$ satisfies the periodic relations
\begin{align}
\widetilde{\psi}_{i}(x,y,\gamma +  2\pi \ell,\eta) &= e^{i(v - 2\pi x)} e^{i\eta} \widetilde{\psi}_{i}(x,y,\gamma,\eta) \\
\widetilde{\psi}_{i}(x,y,\gamma,\eta +  2\pi \ell) &= e^{i(u - 2\pi y)} e^{-i\gamma} \widetilde{\psi}_{i}(x,y,\gamma,\eta)\ . 
\end{align}
If one considers $x,y,u,v$ as parameters, then $(\gamma, \eta) \mapsto \widetilde{\psi}_{i}(x,y,\gamma,\eta)$ is a section of a line vector bundles over $(\gamma, \eta)$ with periodicity $2 \pi \ell$, which we denote by $L(x,y,u,v)$. It is easy to see that all these line vector bundles are isomorphic to one of them, say $L_0 = L(0,0,0,0)$. In particular, they have the same first Chern class. 

For any $i=1, \dots, \ell$, let us introduce the quantity
\begin{equation}
\Omega_i(\gamma, \eta) = \int_{[0,1]^2}dxdy\ \left( \frac{\partial \widetilde{\psi}_i^\ast}{\partial \gamma} \frac{\partial \widetilde{\psi}_i}{\partial \eta} - \frac{\partial \widetilde{\psi}_i^\ast}{\partial \eta} \frac{\partial \widetilde{\psi}_i}{\partial \gamma}\right)\ .
\end{equation}
Using the relations (\ref{vaction}) and the partial normalization (\ref{partialnormalizationpsi}), one can show that $\Omega_{i+1}(\gamma, \eta) = \Omega_i(\gamma + 2\pi, \eta)$. Then we get
\begin{equation}
\label{omegagamma}
\int_{[2m\pi, 2(m+1)\pi]\times [0,2\pi]}d\gamma d\eta \ \Omega_{i+1}(\gamma, \eta) = \int_{[2(m+1)\pi, 2(m+2)\pi]\times [0,2\pi]}d\gamma d\eta \ \Omega_{i}(\gamma, \eta)\ .
\end{equation}
Similarly, using (\ref{uaction}) and (\ref{partialnormalizationpsi}), one can show that $\Omega_{i}(\gamma, \eta) = \Omega_i(\gamma, \eta + 2\pi)$ and
\begin{equation}
\label{omegaeta}
\int_{[0, 2\pi]\times [2n \pi,2(n+1) \pi]}d\gamma d\eta \ \Omega_{i}(\gamma, \eta) = \int_{[0, 2\pi]\times [2(n+1) \pi,2(n+2) \pi]}d\gamma d\eta \ \Omega_{i}(\gamma, \eta)\ .
\end{equation}

Assuming the validity of formula (\ref{hallconductivitydegen}), in order to compute the Hall conductivity, one has to evaluate expressions of the form
\begin{equation}
\frac{1}{\ell} \sum_{i=1}^\ell \int_{[0,2\pi]^2} d\gamma d\eta\ \Omega_{i}(\gamma, \eta)\ .
\end{equation}
Taking account of (\ref{omegagamma}) and (\ref{omegaeta}), one get
\begin{align}
\frac{1}{\ell} \sum_{i=1}^\ell \int_{[0,2\pi]^2} d\gamma d\eta\ \Omega_{i}(\gamma, \eta) &=  \frac{1}{\ell} \int_{[0,2\pi\ell] \times [0,2\pi]} d\gamma d\eta\ \Omega_{1}(\gamma, \eta) \nonumber \\
 &= \frac{1}{\ell^2} \sum_{n=0}^{\ell-1} \int_{[0,2\pi\ell] \times [2 n \pi,2(n+1) \pi]} d\gamma d\eta\ \Omega_{1}(\gamma, \eta) \nonumber \\
&= \frac{1}{\ell^2} \int_{[0,2\pi \ell]^2} d\gamma d\eta\ \Omega_{1}(\gamma, \eta)\ .
\end{align}
This last integral can be computed geometrically using the first Chern class of the line vector bundle $L_0$ and results of Appendix~\ref{computechern}, giving $\frac{4 \pi i}{\ell}$. If there were only one sector for the representation of $\widetilde{U}^\frac{1}{\ell}$ and $\widetilde{V}^\frac{1}{\ell}$, the Hall conductivity would then be
\begin{equation}
\label{sigmaxyNB}
\sigma_{xy} = \frac{N^2}{i(2\pi)^2} \frac{4 \pi i}{\ell} = 2 \frac{N}{B}
\end{equation}
where we use $2 \pi \ell = NB$. When there is many sectors in the representation of $\widetilde{U}^\frac{1}{\ell}$ and $\widetilde{V}^\frac{1}{\ell}$, assuming yet its validity, the formula (\ref{hallconductivitydegen}) giving the Hall conductivity decomposes into an average\footnote{Because the parameters $u$ and $v$ may be continuous, this average can take the form of an integral.} over the contribution of each sector, with equal weight. Because these contributions are the same, one obtains again (\ref{sigmaxyNB}).

In our units, the elementary flux $\phi_0$ is $2\pi$ and $e^2/h$ is $\frac{1}{2\pi}$, so that the number of magnetic fluxes though the sample is $N_\phi = \frac{B}{2\pi}$. Then, $\sigma_{xy}$ has finally the expression
\begin{equation}
\sigma_{xy} = 2 \frac{e^2}{h} \frac{N}{N_\phi}\ .
\end{equation}

In this computation, we used only the dependence of the wave functions in the global variables $x$ and $y$. As noticed previously in the text and in Appendix~\ref{kuboformula}, the structure of the Kubo formula permits one to avoid explicit reference to the $\cH_I$ part of the Hilbert space, and only general properties on the part of the wave functions in this Hilbert space are used, for instance their orthonormalisations.

We now make some comments about the different geometrical structures appearing in this framework. The first line vector bundle which is involved has $(x,y) \in [0,1]^2$ as base space. Its first Chern class is $BN/(2\pi) = \ell$. Notice that the magnetic field, taken alone, does not need to be of the form $2 \pi \ell$, which appears to be different from other mathematical models that already appeared in the literature. This reflects the fact that we used global variables. This line vector bundle has no physical existence, and it is only a mathematical tool used to compute the conductivity. The second line vector bundle has $(\gamma, \eta) \in [0,2 \pi \ell]^2$ for base space. Its first Chern class is $2 \ell$. Its appearance and its structure are strongly related to the first line vector bundle through the unitary symmetry present in the model. Indeed, this symmetry connects directly some translations in the first base space with some in the second base space. As seen in the computation, the degeneracy associated to each sector of the representation of the unitary symmetries $\widetilde{U}^\frac{1}{\ell}$ and $\widetilde{V}^\frac{1}{\ell}$ is related to the line vector bundle over $(\gamma, \eta) \in [0,2 \pi \ell]^2$ when one assembles these functions into a global section of this bundle. Similar geometrical structures have been considered previously in the literature (see e.g. \cite{WenNiu:90}). Notice finally that our computation fails when $NB \neq 2 \pi \ell$. In this case, the first line vector bundle does not exist, and the family of unitary symmetries cannot be implemented easily.

\section{Comments and conclusion}

\label{commentsandconclusion}

In this paper we have presented a new formalism which permits one to deal with some quantum mechanical aspects of the QHE. This formalism has permitted us to compute the Hall conductivity, provided some physical assumptions are made about the applicability of the Kubo formula and its variant in presence of degeneracy. This formalism has been introduced using physical motivations. Its main caracteristic is that it take into account all the possible boundary conditions compatible with the experimental situation. This is mathematically translated into the fact that the representation of the algebra (\ref{algebra}) is \emph{reducible}. This reducibility is the key ingredient for the computation of the Hall conductivity, where this representation space is decomposed (continuously) into representations of some unitary symmetries. These symmetries help us to characterize the line vector bundle whose first Chern class appears in the Kubo formula, because they connect a line vector bundle over the space $(x,y)$ and the latter line vector bundle over the space $(\gamma, \eta)$.

Let us comment now the results of the computation. The experiment predicts a value for the Hall conductivity which is $e^2/h$ times a fraction. This fraction is usually identified with the theoretical filling factor defined by $\nu = \frac{N}{N_\phi}$. Actually, this definition of the filling factor makes explicit reference to the Landau model on the whole plane $\gR^2$, in which the number of states that the system can occupy is known. Here it is not the case. Our result agrees however with the experiment, since it predicts a Hall conductivity $e^2/h$ times a fraction, and the dependence of this fraction with respect to $B$ and $N$ is the expected one. At present time, we do not have any satisfactory explanation for the factor $2$ appearing here in the square shaped sample case.

Notice that in our formalism and computation, and actually in any model based on topological considerations, a possible difference between the odd and even denominators of the filling factor has nowhere emerged. This implies that in order to explain all the experimental observations, one has to make some refinements to the model. A plausible situation is that the term of the Hamiltonian for the ``internal'' variables may play a crucial part here. 

Our model does not explain, as well, the observed stability of the plateaus. In order to get some information in this direction, one would have to compute the Hall conductivity for $B$ such that  $NB \neq 2 \pi \ell$, but with $NB$ \emph{near} $2 \pi \ell$. In this case, as already noticed, our computation fails, because we do not have a line vector bundle on the space $(x,y)$ anymore.

\appendix

\renewcommand{\theequation}{\thesection.\arabic{equation}}

\section{Expression of the Kubo formula}
\label{kuboformula}
\setcounter{equation}{0}

The Kubo formula gives us the Hall conductivity in the form
\begin{equation}
\sigma_{xy} = i \sum_{n > 0} \frac{(v_y)_{0n}(v_x)_{n0} - (v_x)_{0n}(v_y)_{n0}}{(E_0 - E_n)^2}
\end{equation}
where $0$ denotes the ground state and $n$ the excited states of the Hamiltonian for the $N$ electrons in the absence of electric field, $(v_x)_{0n}$ is defined with the total velocity operator $v_x = \frac{1}{m}P_x$ as $(v_x)_{0n} = \langle 0 \vert v_x \vert n \rangle$, and the values $e=1$ and $\hbar=1$ have been assumed. The usual procedure to compute the mean value of the velocity is to derive the Hamiltonian by  parameters which appear as an additional terms in the momentum operators, and take the mean value of these derivatives \cite{NiuThouWu:85, Kohm:85, AvroSeil:85, NiuThou:87}. 

In our case, notice that this computation concerns only the global variables of the Hamiltonian. It can be easily done if one performs the gauge transformation (\ref{gaugetransformation}) and uses the parameters $\gamma,\eta$ to derive the new Hamiltonian $\widetilde{H}_0$ obtained by this gauge transformation. We assume that the remaining part of the wave function which depends only on the internal variables is normalized to $1$. We would like to emphasizes that there is here an extra factor $N^2$ which comes from the definition of the global variable; indeed, one has $v_x = N \frac{\partial \widetilde{H}_0}{\partial \gamma}$. All computations done, following \cite{NiuThouWu:85, Kohm:85, NiuThou:87}, one obtains
\begin{equation}
\label{hallconductivity}
\sigma_{xy} = \frac{N^2}{i(2\pi)^2} \int_{[0,2\pi]^2} d\gamma d\eta \int_{[0,1]^2} dx dy \left( \frac{\partial \widetilde{\psi}^\ast}{\partial \gamma} \frac{\partial \widetilde{\psi}}{\partial \eta} - \frac{\partial \widetilde{\psi}^\ast}{\partial \eta} \frac{\partial \widetilde{\psi}}{\partial \gamma}\right)\ .
\end{equation}
The integral over $(\gamma, \eta)\in [0,2\pi]^2$ appearing in this formula is a part of the scalar product in $\cH_0$. The extra $1/(2\pi)^2$ factor comes from the normalization of this scalar product. Indeed, if we use the partial normalization (\ref{partialnormalizationpsi}) (see Appendix~\ref{computechern} for the justification), which can be expressed on $\widetilde{\psi}$ as
\begin{equation}
\label{partialnormalization}
\int_{[0,1]^2} dx dy (\widetilde{\psi}^\ast \widetilde{\psi})(x,y,\gamma,\eta) = 1
\end{equation}
for all $\gamma,\eta$, then the factor $1/(2\pi)^2$ comes from the scalar product
\begin{equation}
\frac{1}{(2\pi)^2} \int_{[0,2\pi]^2} d\gamma d\eta \int_{[0,1]^2} dx dy (\widetilde{\psi}^\ast \widetilde{\psi})(x,y,\gamma,\eta) = 1\ .
\end{equation}
An other way to look at this integration is to consider it, as in \cite{NiuThouWu:85}, as an average over all the possible values for $(\gamma, \eta)\in [0,2\pi]^2$. Physically, this means that we are taking account of all the non equivalent representations. 

The expression
\begin{equation}
\label{factorchern}
\frac{1}{2\pi i} \int_{[0,2\pi]^2} d\gamma d\eta \int_{[0,1]^2} dx dy \left( \frac{\partial \widetilde{\psi}^\ast}{\partial \gamma} \frac{\partial \widetilde{\psi}}{\partial \eta} - \frac{\partial \widetilde{\psi}^\ast}{\partial \eta} \frac{\partial \widetilde{\psi}}{\partial \gamma}\right)
\end{equation}
is very similar to the first Chern class of a line vector bundle over a $2$-dimensional torus $(\gamma, \eta)\in [0,2\pi]^2$, expressed here as the integral of the curvature of a ``gauge potential''
\begin{equation}
\label{gaugepotentialKubo}
A = \int_{[0,1]^2} dx dy \left( \widetilde{\psi}^\ast \frac{\partial \widetilde{\psi}}{\partial \gamma} d\gamma + \widetilde{\psi}^\ast \frac{\partial \widetilde{\psi}}{\partial \eta} d\eta \right)\ .
\end{equation}
Actually, we are not sure at all to really have here such a line vector bundle with the torus $(\gamma, \eta)\in [0,2\pi]^2$ as base manifold. Nevertheless, as explained in the text, in some cases, the computation of this integral can be done using a geometric identification of that kind. Such an interpretation of the Kubo formula has been explored first in \cite{AvroSeilSimo:83} in the context of the TKNdN theory \cite{ThouKohmNighNijs:82}.

In case there is a degeneracy in the eigenfunctions of the Hamiltonian, the equation (\ref{hallconductivity}) has to be modified. If there is no coupling between different states in the same eigensubspace, then the Hall conductivity is given by (see~\cite{NiuThouWu:85})
\begin{equation}
\label{hallconductivitydegen}
\sigma_{xy} = \frac{N^2 }{i(2\pi)^2} \frac{1}{D} \sum_{k=1}^D \int_{[0,2\pi]^2} d\gamma d\eta \int_{[0,1]^2} dx dy \left( \frac{\partial \widetilde{\psi}_k^\ast}{\partial \gamma} \frac{\partial \widetilde{\psi}_k}{\partial \eta} - \frac{\partial \widetilde{\psi}_k^\ast}{\partial \eta} \frac{\partial \widetilde{\psi}_k}{\partial \gamma}\right)
\end{equation}
where $D$ is the degree of the degeneracy and the functions $\widetilde{\psi}_k$ constitute an orthogonal basis of the eigensubspace.

\section{Computation of the first Chern class}
\label{computechern}
\setcounter{equation}{0}

In this appendix we give some details on the computation of the first Chern class of a complex line vector bundle over a two dimensional torus $(\gamma,\eta) \in [0, 2 \pi \ell]^2$ for an integer $\ell$.

As a starting point, we take a section $f(\gamma,\eta)$ of the line vector bundle, satisfying the following relations on the boundary of the square $[0, 2 \pi \ell]^2$:
\begin{equation}
f(\gamma + 2\pi\ell, \eta) = e^{i \eta}f(\gamma,\eta) \quad \text{ and }\quad
f(\gamma, \eta + 2\pi\ell) = e^{-i \gamma}f(\gamma,\eta)\ .
\end{equation}
We suppose that $f^\ast(\gamma,\eta) f(\gamma,\eta)=1$ for all $\gamma,\eta$ (this correspond to (\ref{partialnormalization}) in our framework). The connection (gauge potential) we take is of the form (see (\ref{gaugepotentialKubo}))
\begin{equation}
A(\gamma,\eta) = f^\ast(\gamma,\eta) \frac{\partial f}{\partial \gamma}(\gamma,\eta) d\gamma + f^\ast(\gamma,\eta) \frac{\partial f}{\partial \eta}(\gamma,\eta) d\eta = A_\gamma(\gamma,\eta) d\gamma + A_\eta(\gamma,\eta) d\eta
\end{equation}
the curvature of which is
\begin{equation}
dA = \left( \frac{\partial f}{\partial \gamma}^\ast \frac{\partial f}{\partial \eta} - \frac{\partial f}{\partial \eta}^\ast \frac{\partial f}{\partial \gamma}\right) d\gamma \wedge d\eta\ .
\end{equation}
The boundary relations on $f$ mean that $A$ has different expressions if $(\gamma,\eta) \in [0, 2 \pi \ell[^2$, $\gamma \geq 2\pi\ell$ or $\eta \geq 2\pi\ell$. Specifically, one has
\begin{subequations}
\begin{align}
A(\gamma, \eta + 2\pi\ell) &= A(\gamma,\eta) -i d\gamma \\
A(\gamma + 2\pi\ell, \eta) &= A(\gamma,\eta) +i d\eta\ .
\end{align}
\end{subequations}
They are just the ordinary gauge transformations of a connection when one looks at it over two different open sets of the base manifold trivializing the line vector bundle.

The first Chern class is given by
\begin{equation}
n = \frac{1}{2\pi i} \int_{[0, 2 \pi \ell]^2} dA\ .
\end{equation}
It is well known that this is an integer. Using Stokes' theorem, one has
\begin{equation}
n = \frac{1}{2\pi i} \left( \int_0^{2\pi\ell} \left( A_\gamma(\gamma,0) - A_\gamma(\gamma,2\pi\ell)  \right) d\gamma + \int_0^{2\pi\ell} \left( A_\eta(2\pi\ell, \eta) - A_\eta(0,\eta)  \right) d\eta \right)\ .
\end{equation}
From the gauge transformation undergone by $A$ at the boundaries, one gets the final result
\begin{equation}
n = \frac{1}{2\pi i} \left( \int_0^{2\pi\ell} i d\gamma + \int_0^{2\pi\ell} i d\eta \right) = 2 \ell\ .
\end{equation}

\end{document}